\documentclass[twocolumn,prl,a4paper,aps,showpacs]{revtex4}
\usepackage{graphicx}
\usepackage{dcolumn}
\usepackage{bm}

\begin{document}

\title{Magnetic long-range order induced by quantum relaxation in single-molecule magnets}

\author{M. Evangelisti$^{1}$, F. Luis$^{2}$, F. L. Mettes$^{1}$, N. Aliaga$^{3,*}$, G. Arom\'i$^{4}$, J. J. Alonso$^{5}$, G. Christou$^{3}$, and L. J. de Jongh$^{1}$}

\affiliation{$^{1}$ Kamerlingh Onnes Laboratory, Leiden University, 2300 RA Leiden, The
Netherlands\\ $^{2}$ ICMA, CSIC and Universidad de Zaragoza, 50009 Zaragoza, Spain\\ $^{3}$
Department of Chemistry, University of Florida, FL 32611 Gainesville, USA\\ $^{4}$ Departament de
Qu\'imica Inorg\`{a}nica, Universitat de Barcelona, 08028 Barcelona, Spain\\ $^{5}$ Departamento
de F\'isica Aplicada I, Universidad de M\'alaga, 29071 M\'alaga, Spain}
\date{\today}

\begin{abstract}
Can magnetic interactions between single-molecule magnets (SMMs) in a crystal establish long-range
magnetic order at low temperatures deep in the quantum regime, where the only electron
spin-fluctuations are due to incoherent magnetic quantum tunneling (MQT)? Put inversely: can MQT
provide the temperature dependent fluctuations needed to destroy the ordered state above some
finite $T_{c}$, although it should basically itself be a $T$-independent process? Our experiments
on two novel Mn$_{4}$ SMMs provide a positive answer to the above, showing at the same time that
MQT in the SMMs has to involve spin-lattice coupling at a relaxation rate equaling that predicted
and observed recently for nuclear spin-mediated quantum relaxation.
\end{abstract}

\pacs{75.40.Cx, 75.45.+j, 75.50.Xx}

\maketitle

Despite the large number of studies on magnetic quantum tunneling (MQT) in molecular crystals of
single-molecule magnets (SMMs)~\cite{dante}, the question whether it is able to bring the spin
system into thermal equilibrium with the lattice, remains unsolved. Prokof'ev and
Stamp~\cite{fe8th} have suggested that interaction with rapidly fluctuating hyperfine fields can
bring a significant number of electron spins into resonance. Coupling to a nuclear spin bath
indeed allows ground-state tunneling over a range of local bias fields $\xi$ much larger than the
tunnel splitting $\Delta$, whereas, in its absence, tunneling would happen only if $\xi$ is
$\sim\Delta$ or less. Within this theory, magnetic relaxation could thus in principle occur with
no exchange of energy with the phonons of the molecular crystal~\cite{Fernandez03}. Support for
the Prokof'ev/Stamp model came from magnetic relaxation studies on the Fe$_{8}$ SMM~\cite{fe8exp}.
However, these experiments covered only initial stages of the relaxation process, leaving open the
question whether the final state corresponds to a thermal equilibrium. When MQT is combined with
coupling to a heat bath, dipolar couplings between cluster spins can induce long-range magnetic
ordering (LRMO) at sufficiently low temperatures~\cite{julio}. This phenomenon has not been
observed yet for any of the known SMMs relaxing by MQT. Attempts have been made in Fe$_{8}$ and
Mn$_{12}$ in particular, by increasing the tunneling rate by means of an applied transverse
field~\cite{trans}. However, the failure is inherent to the approach, since the applied fields
needed are much higher than the weak interaction energies involved, and will thus destroy the
LRMO.

In this letter, we present zero-field time-dependent specific heat measurements performed on two
novel tetranuclear molecular clusters, both with net cluster spin $S=9/2$, denoted by Mn$_{4}$Cl
and Mn$_{4}$Me, which have similar cluster cores but different ligand molecules. For both, the
cluster spins become blocked along their anisotropy axes for temperatures in the liquid helium
range, and relaxation below $\sim 0.8$~K can thus only proceed by incoherent MQT between the two
lowest lying states $m=\pm S$. For both compounds we prove below that the MQT has to be inelastic.
For Mn$_{4}$Me, the tunneling rates are found sufficiently high to establish, even at zero field,
thermal equilibrium conditions down to the lowest temperatures ($\sim 0.1$~K). Accordingly, the
MQT channel enables the occurrence of LRMO between the cluster spins at $T_{c}=0.21$~K. Comparing
the magnitude of $T_{c}$ with Monte Carlo simulations suggests the coexistence of dipolar and weak
superexchange interactions between clusters. In view of the essential role of the dynamic nuclear
bias in the MQT mechanism, our results call for an extension of the nuclear-spin mediated quantum
relaxation model~\cite{fe8th} to include inelastic processes, where MQT is accompanied by phonon
creation or annihilation.

Analytically pure samples of the compounds Mn$_{4}$O$_{3}$L(dbm)$_{3}$, with L~=~Cl(OAc)$_{3}$ or
[O$_{2}$C(C$_{6}$H$_{4}$-$p$-Me)]$_{4}$, hereafter abbreviated as Mn$_{4}$Cl and Mn$_{4}$Me, were
prepared as described in Ref.~\cite{mag}. All samples were characterized by elemental analysis.
Both molecules possess a distorted cubane core with one Mn$^{4+}$ ion (spin $S=3/2$) and three
Mn$^{3+}$ ions ($S=2$), superexchange coupled via three oxygen ions. The {\it intra}-cluster
exchange couplings have been studied by magnetic susceptibility measurements~\cite{mag}. Below
$T\lesssim 10$~K, the four Mn spins become ordered with a net total spin $S=9/2$ subject to an
uniaxial crystal field~\cite{uniaxial}, the symmetry axis running approximately through the
Mn$^{4+}$ ion and the L-ligand. Whereas Mn$_{4}$Cl has a local virtual $C_{3V}$ symmetry, the more
bulky carboxylate ligand distorts and lowers the symmetry ($C_{S}$) of Mn$_{4}$Me in the crystal
and increases the magnitude of the uniaxial anisotropy~\cite{mag,uniaxial}.

Low-temperature specific heat measurements were performed by a home-made calorimeter using a
thermal relaxation method, see Ref.~\cite{trans}. By varying the thermal resistance of the thermal
link between calorimeter and cold-sink, the characteristic timescale $\tau_{e}$ of the experiment
can be varied. In this way time-dependent specific heat measurements can be exploited to
investigate spin-lattice relaxation when the relaxation time becomes of the order of $\tau_{e}$
($\sim 1-100$~s)~\cite{cm}. The samples consisted of $1-3$~mg of polycrystalline material, mixed
with $2-5$~mg of Apiezon-N grease to ensure good thermal contact. The high-temperature ($2~{\rm
K}<T<300~{\rm K}$) specific heat was measured for a Mn$_{4}$Cl pellet sample of about $40$~mg
using a commercial (PPMS) calorimeter.

The specific heat $C/R$ of Mn$_{4}$Cl is shown in Fig.~1, which combines data obtained for
$\tau_{e}\approx 2$~s, $8$~s, and $300$~s (as estimated at $T=0.4$~K) with the ones obtained with
the high-$T$ calorimeter. Let us first consider data measured for $T>1$~K, where $C$ is
independent of $\tau_{e}$. A $\lambda$-type anomaly is observed at $T\simeq 7$~K, having a
relative height of $2.5$~$R$. Measurements with an applied magnetic field (not presented here)
have shown that this anomaly is insensitive to the field, proving that it has to be associated to
a structural phase transition. Between $1$~K and $7$~K, $C$ is dominated by contributions from the
lattice phonons and from transitions between energy levels of the $S=9/2$ multiplet split by the
crystal field.

Neglecting, in first approximation, the Zeeman splittings $\xi$ due to effective fields
representing the hyperfine interaction and the {\it inter}-cluster magnetic couplings, the
spectrum of energy levels is doubly degenerate at zero-applied-field, so that $C$ only depends on
transitions between levels inside each of the potential wells that are separated by the anisotropy
energy barrier $U$. We shall call this the {\it intra}-well contribution. The associated
multilevel Schottky anomaly is calculated with the eigenvalues of the spin Hamiltonian
\begin{equation}\label{Hamiltonian}
{\cal H}=-DS_{z}^{2}-E(S_{x}^{2}-S_{y}^{2})+A_{4}S_{z}^{4}
\end{equation}
with parameters $D$, $E$ and $A_{4}$ obtained independently from inelastic neutron scattering and
high-frequency EPR measurements~\cite{epr,parameters}. The {\it intra}-well specific heat
decreases exponentially as $T$ decreases and, since the first excited $m=\pm 7/2$ level is about
$(2S-1)D=5.5$~K above the $m=\pm 9/2$ ground-state doublet, it becomes almost negligible when $T
\leq 0.8$~K (Fig. 1). Adding the lattice contribution, calculated with a Debye temperature
$\theta_{D}\simeq 15$~K, to the Schottky accounts well for the experimental data (solid line in
Fig.~1). The lattice specific heat above 10~K appears to be composed of a number of Einstein-type
contributions (not shown).

\begin{figure}[t!]
\centering{\includegraphics[angle=0,width=8cm]{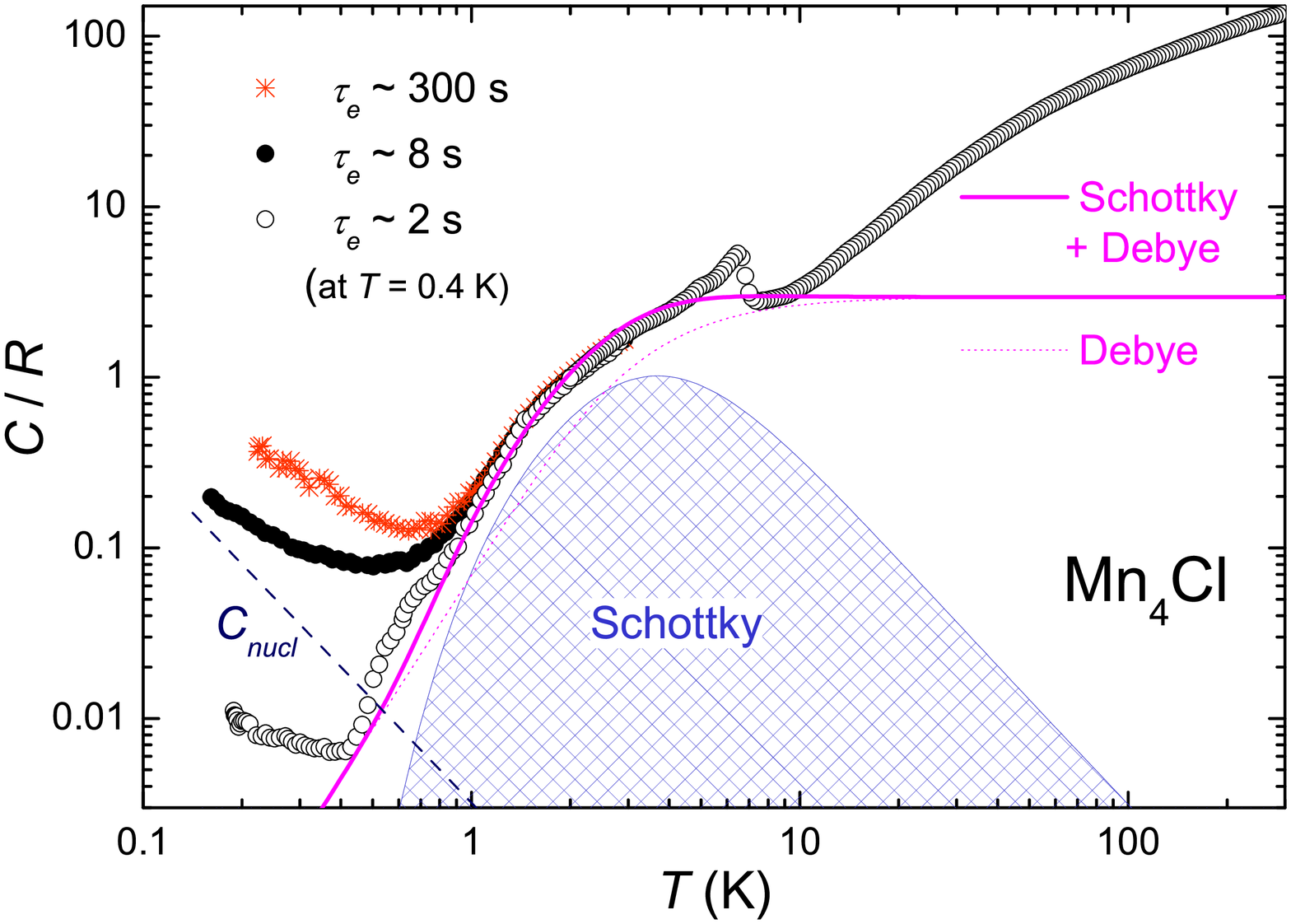}\caption{Temperature dependence of the
specific heat of Mn$_{4}$Cl measured for $\tau_{e}\approx 2$~s, 8~s and 300~s. Solid line, sum of
the Debye term of the lattice (dotted line) plus the Schottky contributions; dashed line,
calculated nuclear contribution $C_{nucl}$.}}
\end{figure}

Below $1$~K, we expect the equilibrium magnetic specific heat ($C_{m}$) to be dominated by two
contributions. The first arises from incoherent MQT events inside the ground-state doublet that is
split by the action of the effective fields arising from hyperfine interactions and {\it
inter}-cluster dipolar coupling. The second is the specific heat $C_{nucl}$ of the nuclear spins
of Mn, whose energy levels are split by the hyperfine interaction with the atomic electron spins.
The dashed line in Fig.~1 represents $C_{nucl}$ calculated with the hyperfine constants
$A_{hf}=7.6$~mK and $A_{hf}=11.4$~mK for, resp., Mn$^{3+}$ and Mn$^{4+}$ ions, obtained from
ESR~\cite{chf}. Experiments performed for the longest $\tau_{e}\approx 300$~s show indeed a large
low-$T$ contribution. By contrast, the specific heat decreases by almost two orders of magnitude
when $\tau_{e}\approx 2$~s, evidencing that $\tau_{e}$ has a large effect in this temperature
range. This shows that the equilibrium between the relative populations of the $+9/2$ and $-9/2$
states cannot be established within $\tau_{e}$ if this is too short. We note that, for the
shortest $\tau_{e}$, the low-$T$ specific heat becomes even smaller than $C_{nucl}$, indicating
that both nuclear and electron spins are out of equilibrium. This is understandable, since the
only channel for the nuclear spins to exchange energy with the lattice is via the electron spins.
The strong connection between nuclear and electron spin-lattice relaxation has earlier been
observed for Mn$_{12}$, Mn$_{6}$ and Fe$_{8}$~\cite{trans,isotropic}.

\begin{figure}[t!]
\centering{\includegraphics[angle=0,width=8cm]{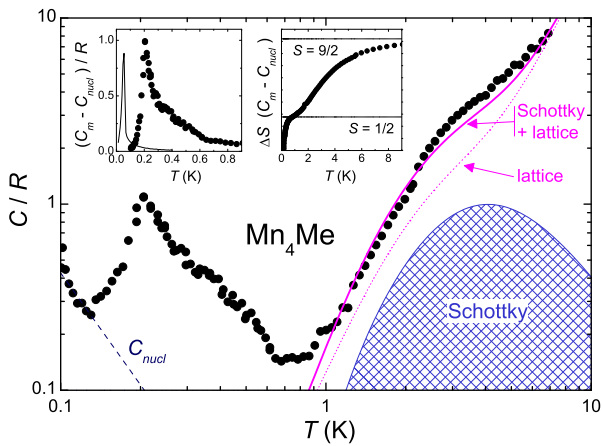}\caption{$T$-dependent specific heat of
Mn$_{4}$Me measured for $\tau_{e}\approx 4$~s. Dashed line, calculated nuclear contribution
$C_{nucl}$; solid line, sum of lattice (dotted line) plus Schottky contributions. Insets:
electronic $C_{m}$ and entropy variation $\Delta S$, after subtraction of $C_{nucl}$. Solid line,
calculated dipolar ordering for the easy axis along the $(110)$ direction (see text); dashed
lines, high-$T$ entropy limits for $S=1/2$ and $S=9/2$.}}
\end{figure}

The experimental $C/R$ of Mn$_{4}$Me, measured for $\tau_{e}\approx 4$~s (as estimated at
$T=0.4$~K), is shown in Fig.~2 together with the calculated contributions of the lattice and the
nuclear spins, and the {\it intra}-well Schottky contribution. The experimental data display a
$\lambda$-anomaly at $T_{c}=0.21$~K that we attribute to the onset of LRMO. The Schottky anomaly
is calculated as for Mn$_{4}$Cl with parameters obtained from high-frequency
EPR~\cite{epr,parameters}. For $T>1$~K, the remaining contribution is given by the lattice and is
well described by the sum of a Debye term for the acoustic low-energy modes plus an Einstein term
for a higher energy mode, with values $\theta_{D}\simeq 12.3$~K and $\theta_{E}\simeq 22$~K for,
resp., the Debye and Einstein temperatures. Below $0.15$~K, the specific heat of Mn$_{4}$Me shows
a clear upturn that can be described by $C_{nucl}/R \simeq 4.27\times 10^{-3}/T^{2}$. This
contribution is well fitted using hyperfine constants $A_{hf}=8.7$~mK and $A_{hf}=13.8$~mK for,
resp., Mn$^{3+}$ and Mn$^{4+}$ ions. These values are of the same order as for Mn$_{4}$Cl and are
close to those reported by Zengh and Dismukes for a natural Mn$_{4}$ cluster~\cite{chf}.

As seen in Fig.~2, the magnetic ordering peak and the calculated Schottky due to the splitting of
the $S=9/2$ multiplet are well separated in temperature. This is confirmed by the analysis of the
temperature dependence of the electronic entropy $\Delta
S(T)=\int^{\infty}_{0}[C_{m}(T)-C_{nucl}(T)]/T{\rm d}T$ (right inset of Fig.~2). As expected, the
total entropy of the electron spins tends to $R\ln(2S+1)$, with $S=9/2$, at high temperatures.
However, for the magnetic ordering region ($T<0.8$~K), $\Delta S$ corresponds to an effective spin
$S=1/2$, as appropriate for a two-level system. This proves that there only the two lowest levels
($m=\pm 9/2$) are populated and contribute to $C_{m}$.

We note that, although the anisotropy barrier $U\simeq DS^{2}-A_{4}S^{4}-|E|S^{2}$ amounts to
$\approx 14$~K for both Mn$_{4}$Me and Mn$_{4}$Cl, due to the lower symmetry of Mn$_{4}$Me the
2nd-order off-diagonal term ($E$) in the Hamiltonian of Eq.~(\ref{Hamiltonian}), and thus also
$\Delta$ of the ground-state, will be much larger for this compound. This is confirmed by
high-frequency EPR experiments that give $E/D \simeq 0.21$, i.e., nearly $5$ times larger than for
Mn$_{4}$Cl \cite{epr,parameters}. We observe that this difference has a very large influence on
the spin-lattice relaxation. For similar $\tau_{e}$ values, the electron spins of the Mn$_{4}$Cl
molecule go off equilibrium below $T=0.8$~K (Fig.~1), whereas for Mn$_{4}$Me we observe thermal
equilibrium for electron and nuclear spins down to the lowest temperature. We conclude this from
the fact that ({\it i}) the total electronic entropy contribution equals the expected limit for
$S=9/2$; ({\it ii}) the remaining specific heat below 0.15~K agrees well with the calculated
$C_{nucl}$.

As commonly found in molecular clusters, the magnetic core of Mn$_{4}$Me is surrounded by a shell
of non-magnetic ligand molecules. It follows that, {\it inter}-cluster superexchange interactions
are very weak, leaving the {\it inter}-cluster dipolar coupling responsible for magnetic
ordering~\cite{isotropic}. To check this for Mn$_{4}$Me, we have performed Monte Carlo
simulations, as described in Ref.~\cite{julio}, for a $S=9/2$ Ising model of magnetic dipoles
regularly arranged on the Mn$_{4}$Me lattice. We have repeated our calculations for several
orientations of the molecular easy axis (an example is given in the left inset of Fig.~2 for the
easy axis along the $(110)$ direction). The calculated $T_{c}$'s are always much smaller than the
experimental value. Consequently, the Mn$_{4}$Me molecules are also coupled by weak superexchange
interactions. Indeed, by adding an {\it inter}-cluster nearest-neighbor exchange interaction
$|J|/k_{B}\simeq 0.14$~K to our dipolar calculations, we reproduce the experimental $T_{c}$
value~\cite{juanjo}.

In order to estimate the spin-lattice relaxation rate $\Gamma$ for Mn$_{4}$Cl at low temperatures,
we have used the relation for the time-dependent specific heat
$C_{m}(t)=C_{0}+(C_{eq}-C_{0})\left[1-e^{-\Gamma\tau_{e}}\right]$, where $C_{0}$ and $C_{eq}$ are
the adiabatic and equilibrium limits of the specific heat, respectively~\cite{cm}. For the
electron spins, $C_{0}$ is to good approximation given by the {\it intra}-well Schottky specific
heat, whereas the ``slow'' specific heat at equilibrium corresponds to excitations involving
transitions between the two wells. We have fitted, thus, the $C_{m}(\tau_{e})$ data of Mn$_{4}$Cl,
taking at each temperature for $C_{eq}$ the $C_{m}(T)$ measured for Mn$_{4}$Me (in the range
$T>T_{c}$). We show $\Gamma(T)$ in Fig.~3, together with data from ac-susceptibility and
(short-time) magnetic relaxation experiments~\cite{xmn4}. For $T>1.7$ K, $\Gamma$ follows the
Arrhenius law, with $U\simeq 13.5$~K and $\tau_{0}\approx 1.4\times 10^{-7}$~s. For $T\lesssim
0.8$~K, $\Gamma$ deviates from this thermal-activation law, in remarkable agreement with the
magnetic relaxation data. Its weak temperature dependence confirms that relaxation to thermal
equilibrium is dominated by direct MQT transitions within the ground-state doublet.

\begin{figure}[t!]
\centering{\includegraphics[angle=0,width=8cm]{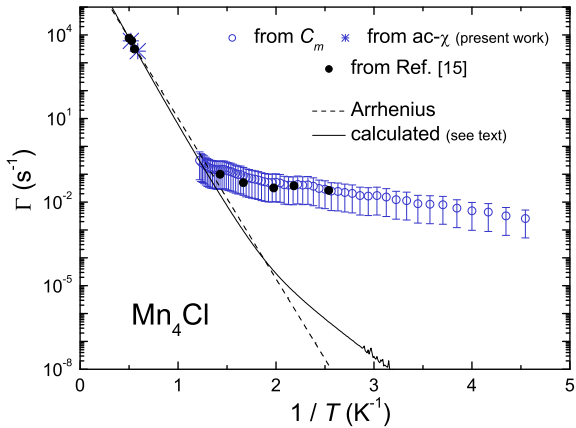}\caption{Spin-lattice relaxation rate of
Mn$_{4}$Cl: ($\ast$) and ($\bullet$)~\cite{xmn4} obtained from ac-susceptibility and magnetic
relaxation data; ($\circ$) obtained by fitting the $C_{m}(\tau_{e})$ data of Fig.~1 (see text).
Dashed line is fit of high-$T$ data to Arrhenius law; solid line is calculated for magnetic fields
$B_{x}=150$~G and $B_{z}=350$~G (see text).}}.
\end{figure}

Summing up, we have proven that for both compounds thermal contact between spins and lattice is
established by MQT fluctuations. For Mn$_{4}$Me, the associated rate is even fast enough to
produce thermal equilibrium down to the lowest temperature and thus enable LRMO. This implies
$\Gamma\gtrsim 1/\tau_{e}\simeq 1$~s$^{-1}$. For Mn$_{4}$Cl, we find a lower rate
($10^{-1}-10^{-3}$~s$^{-1}$) for $T\lesssim 0.8$~K. It is of interest to compare the experimental
$\Gamma$ for Mn$_{4}$Cl with predictions from conventional models for spin-lattice relaxation,
assuming that the $m=\pm 9/2$ energy levels of the cluster spins are time-independent. We have
simulated the effect of {\it inter}-cluster dipolar coupling and hyperfine interactions by
introducing static magnetic fields $B_{x}=150$~G and $B_{z}=350$~G~\cite{fields}. The presence of
these Zeeman terms is essential, otherwise tunneling would be forbidden for half-integer spin
$S=9/2$~\cite{halfodd}. We calculated $\Gamma$ by solving a master equation, including {\it
intra}- as well as {\it inter}-well transitions, induced by phonons only, between exact
eigenstates of the spin Hamiltonian of Eq.~(\ref{Hamiltonian}), as discussed in Ref.~\cite{cm} and
recently applied to the analysis of $C_{m}$ of Fe$_{8}$ and Mn$_{12}$ clusters~\cite{trans}. The
result (solid line in Fig.~3) agrees well with the activated behavior observed at high-$T$, but
fails to account for $\Gamma$ measured below $1$~K by six orders of magnitude! This large
discrepancy can not be ascribed to errors in the estimated elastic properties of the lattice. Both
the pre-factor of the Arrhenius law and the measured value of $\theta_{D}$ give a value of $c_{s}
\simeq 5(1)\times 10^{2}$ m/s for the speed of sound. By contrast, $\Gamma$ observed at low-$T$
would require $c_{s}$ and $\theta_{D}$ to be $15$ times smaller. This would give rise to a large
lattice specific heat well below $1$ K, which is not observed.

It appears therefore that extension to {\it dynamic} hyperfine fields acting on the cluster-spin
levels, as proposed in Ref.~\cite{fe8th}, is indeed a necessary prerequisite for any model for the
MQT of SMMs. Such dynamic bias fields will sweep the tunneling levels with respect to one another,
thereby enabling incoherent Landau-Zener type tunneling events. The model predicts quantum
relaxation rates agreeing with experiments~\cite{fe8exp}, but so far the relaxation of the cluster
spins was thought to occur solely/primarily to the nuclear spin bath, relaxation to phonons was
only expected at much longer time-scales. Our specific heat experiments, in which obviously the
heat is transferred to the spins {\it via} the lattice, clearly demonstrate that in fact
spin-lattice relaxation {\it has to be} involved and at much the same fast rates! Since
application of `conventional' models for spin-lattice relaxation leads to rates orders of
magnitude too low, our data call for an extension of the Prokof'ev/Stamp model in which nuclear
spin-mediated MQT events are combined with creation or annihilation of phonons.

The authors are indebted to J.F. Fern\'andez and P.C.E. Stamp for enlightening discussions, and
R.S. Edwards and S. Hill for HFEPR experiments on Mn$_{4}$Me. This work is part of the research
program of the Stichting voor Fundamenteel Onderzoek der Materie (FOM).


\begin{thebibliography}{99}

\bibitem[*]{byline} Present address: Max-Planck-Institut f\"ur Strahlenchemie,
45470 M\"ulheim an der Ruhr, Germany.
\bibitem{dante} See, for instance, D. Gatteschi and R. Sessoli, Angew.
Chem. Int. Edit. {\bf 42}, 268 (2003).
\bibitem{fe8th} N.V. Prokof'ev and P.C.E. Stamp, J. Low Temp. Phys. {\bf 104},
143 (1996); Phys. Rev. Lett. {\bf 80}, 5794 (1998); Rep. Prog. Phys. {\bf 63}, 669 (2000).
\bibitem{Fernandez03} J.F. Fern\'andez and J.J. Alonso, Phys. Rev. Lett. {\bf
91}, 047202 (2003).
\bibitem{fe8exp} W. Wernsdorfer {\it et al.}, Phys. Rev. Lett. {\bf 84}, 2965 (2000).
\bibitem{julio} J.F. Fern\'andez and J.J. Alonso, Phys. Rev. B {\bf 62}, 53 (2000);
J.F. Fern\'andez, {\it ibid.} {\bf 66}, 064423 (2002).
\bibitem{trans} F. Luis {\it et al.}, Phys. Rev. Lett. {\bf 85}, 4377 (2000);
F.L. Mettes, F. Luis, and L.J. de Jongh, Phys. Rev. B {\bf 64}, 174411 (2001).
\bibitem{mag} H. Andres {\it et al.}, J. Am. Chem. Soc. {\bf 50}, 12469 (2000);
G. Arom\'i {\it et al.}, Inorg. Chem. {\bf 41}, 805 (2002); N.
Aliaga, K. Folting, D.N. Hendrickson, and G. Christou, Polyhedron
{\bf 20}, 1273 (2001).
\bibitem{uniaxial} S. Wang {\it et al.}, Inorg. Chem. {\bf 35}, 7578 (1996).
\bibitem{cm} J.F. Fern\'andez, F. Luis, and J. Bartolom\'e, Phys. Rev. Lett. {\bf 80},
5659 (1998).
\bibitem{epr} R.S. Edwards and S. Hill (private communication).
\bibitem{parameters} The parameters used in our calculations are $D=0.69$~K,
$E=-3.15\times 10^{-2}$~K, $A_{4}=-3.25\times 10^{-3}$~K for Mn$_{4}$Cl (from Ref.~\cite{mag}),
and $D=0.76$~K, $E=0.15$~K, $A_{4}=-4.3\times 10^{-3}$~K for Mn$_{4}$Me (from HFEPR~\cite{epr}).
\bibitem{chf} M. Zheng and G.C. Dismukes, Inorg. Chem. {\bf 35}, 3307 (1996).
\bibitem{isotropic} A. Bino {\it et al.}, Science {\bf 241}, 1479 (1988); A. Morello
{\it et al.}, Phys. Rev. Lett. {\bf 90}, 017206 (2003).
\bibitem{juanjo} For the exchange interaction between the magnetic dipoles, we consider
$\mathcal{H}_{ex}=2J\sum_{i\neq j}~S_{z}^{ (i)}S_{z}^{ (j)}$. Taking $S=1/2$ and $J\neq 0$ only
for nearest dipoles ($z=5$ in Mn$_{4}$Me), we calculate $T_{c}\simeq 0.2$~K for $|J|/k_{B}\simeq
0.14$~K, irrespective of the sign of $J$.
\bibitem{xmn4} S.M.J. Aubin {\it et al.}, J. Am. Chem. Soc. {\bf 120}, 49981 (1998).
\bibitem{fields} Interaction with nuclear spins induces a distribution of bias $\xi$
of width $E_{0}\simeq N^{1/2}\hbar\omega_{0}/2$, where $N$ is the number of nuclear spins and
$\hbar\omega_{0}$ is the average hyperfine splitting~\cite{fe8th}. Off-diagonal terms of the
hyperfine interaction are of the same order of magnitude~\cite{chf}. In addition, {\it
inter}-cluster dipolar energies $E_{int}$ further broaden the distribution of bias. We roughly
estimate $E_{0}\approx 9\times 10^{-2}$~K and $E_{int}\approx k_{\rm{B}}T_{c}=0.2$~K, equivalent
to $B_{x}\simeq 150$~G and $B_{z}\simeq 350$~G, respectively.
\bibitem{halfodd} D. Loss, D.P. DiVincenzo, and G. Grinstein, Phys. Rev. Lett. {\bf 69},
3232 (1992); J. von Delft and C.L. Henley, {\it ibid.} {\bf 69}, 3236 (1992).
\end{thebibliography}
\end{document}